%% file: unitri7.tex
\documentclass[12pt]{article}

\usepackage{amsmath,amssymb,amsthm}
\usepackage{latexsym,color,graphicx}

\def\eps{{\varepsilon}}
\def\A{{\cal A}}

\def\reals{{\mathbb R}}

\newtheorem{theorem}{Theorem}[section]
\newtheorem{lemma}[theorem]{Lemma}

\newcommand\remove[1]{}

\begin{document}

\title{An Improved Bound on the Number of Unit Area Triangles\thanks{%
  Work on this paper was supported by NSF Grants CCF-05-14079 and
  CCF-08-30272, 
  by a grant from the U.S.-Israeli Binational Science Foundation, 
  by grant 155/05 from the Israel Science Fund and by the
  Hermann Minkowski--MINERVA Center for Geometry at Tel Aviv University.} }
\author{
    Roel Apfelbaum\thanks{%
    School of Computer Science, Tel Aviv University,
    Tel Aviv 69978, Israel;
    \texttt{roel6@hotmail.com}.}
    \and
  Micha Sharir\thanks{%
    School of Computer Science, Tel Aviv University, Tel~Aviv 69978,
    Israel; and Courant Institute of Mathematical Sciences, New York
    University, New York, NY~~10012,~USA. E-mail:
    \texttt{michas@post.tau.ac.il}} }

\maketitle

\begin{abstract}
We show that the number of unit-area triangles determined by a set of
$n$ points in the plane is $O(n^{9/4+\eps})$, for any $\eps>0$,
improving the recent bound $O(n^{44/19})$ of Dumitrescu et al.
\end{abstract}

\section{Introduction}
\label{sec:intro}

In 1967, A.~Oppenheim (see \cite{EP95}) asked the following question:
Given $n$ points in the plane and $A>0$, how many triangles spanned
by the points can have area $A$? By applying a scaling transformation,
one may assume $A=1$ and count the triangles of {\em unit} area.
Erd\H{o}s and Purdy~\cite{EP71} showed that a $\sqrt{\log n}\times
(n/\sqrt{\log n})$ section of the integer lattice determines
$\Omega(n^2 \log\log{n})$ triangles of the same area. They also showed
that the maximum number of such triangles is at most $O(n^{5/2})$. In
1992, Pach and Sharir~\cite{PS92} improved the bound to
$O(n^{7/3})$, using the Szemer\'edi-Trotter theorem~\cite{ST83} 
on the number of point-line incidences. Recently,
Dumitrescu et al.~\cite{DST08} have further improved the upper bound 
to $O(n^{44/19})=O(n^{2.3158})$, by estimating the number of incidences
between the given points and a 4-parameter family of quadratic curves. 

In this paper we further improve the bound to $O(n^{9/4+\eps})$, for
any $\eps>0$. Our proof borrows some ideas from \cite{DST08}, but
works them into a different approach, which reduces the problem to 
bounding the number of incidences between points and certain kind 
of surfaces in three dimensions.

\section{Unit-area triangles in the plane}
\label{sec:unit2}

To simplify the notation, we write $O^*(f(n))$ for an upper bound of
the form $C_\eps f(n)\cdot n^\eps$, which holds for any $\eps>0$,
where the constant of proportionality $C_\eps$ depends on $\eps$.

\begin{theorem}\label{thm:unit2}
The number of unit-area triangles spanned by $n$ points in the plane
is $O^*(n^{9/4})$.
\end{theorem}
\noindent{\bf Proof.}
We begin by borrowing some notation and preliminary ideas from
\cite{DST08}.
Let $S$ be the given set of $n$ points in the plane. Consider a triangle
$\Delta=\Delta{abc}$ spanned by $S$. We call the three lines containing the
three sides of $\Delta{abc}$, {\em base lines} of $\Delta$, and the
three lines parallel to the base lines and incident to the respective
third vertices, {\em top lines} of $\Delta$.

For a parameter $k$, $1\leq k\leq \sqrt{n}$, to be optimized
later, call a line $\ell$ {\em $k$-rich} (resp., {\em $k$-poor}) if
$\ell$ contains at least $k$ (resp., fewer than $k$) points of $S$.
Call a triangle $\Delta{abc}$ {\em $k$-rich} if each of its three top
lines is $k$-rich; otherwise $\Delta$ is {\em $k$-poor}. 

We first observe that the number of $k$-poor unit-area triangles
spanned by $S$ is $O(n^2k)$. Indeed, assign a $k$-poor unit-area
triangle $\Delta{abc}$ whose top line through $c$, say, is $k$-poor
to the opposite base $ab$. Then all the triangles assigned to a base
$ab$ are such that their third vertex lies on one of the two lines
parallel to $ab$ at distance $2/|ab|$, where that line contains
fewer than $k$ points of $S$. Hence, a base $ab$ can be assigned at
most $2k$ triangles, and the bound follows.

So far, the analysis follows that of \cite{DST08}.
We now focus the analysis on the set of $k$-rich unit-area triangles
spanned by $S$, and use a different approach.

Let $L$ denote the set of $k$-rich lines, and let $Q$ denote the set 
of all pairs
$$\{ (\ell,p) \mid \ell\in L,\, p\in S\cap\ell \}.$$
By the Szemer\'edi-Trotter theorem~\cite{ST83}, we have,
for any $k\leq \sqrt{n}$, $m:=|L|=O(n^2/k^3)$, and
$N:=|Q|=O(n^2/k^2)$.

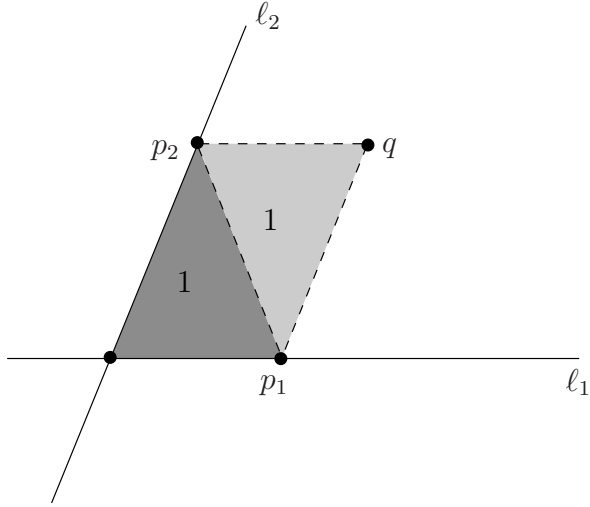
\begin{figure}[htb]
\begin{center}
\input{qmatch.pstex_t}
\caption{The ordered pair $((\ell_1,p_1),(\ell_2,p_2))$ is a matching
         pair of elements of $Q$.}
\label{qmatch}
\end{center}
\end{figure}

A pair $(\ell_1,p_1)$, $(\ell_2,p_2)$ of elements of $Q$ is said
to {\em match} if the triangle with vertices $p_1$, $p_2$,
$\ell_1\cap\ell_2$ has area $1$; see Figure~\ref{qmatch}.

To upper bound the number of unit-area triangles, all of whose
three top lines are $k$-rich, it suffices to bound the number of
matching pairs in $Q$. Indeed, given such a unit-area triangle 
$\Delta p_1p_2q$, let $\ell_1$ (resp., $\ell_2$) be the top line of 
$\Delta p_1p_2q$ through $p_1$ (resp., through $p_2$). 
Then $(\ell_1,p_1)$ and $(\ell_2,p_2)$ form a
matching pair in $Q$, by definition (again, see Figure \ref{qmatch}).
Conversely, a matching pair
$(\ell_1,p_1)$, $(\ell_2,p_2)$ determines at most one unit-area
triangle $p_1p_2q$, where $q$ is the intersection point of the line
through $p_1$ parallel to $\ell_2$ and the line through $p_2$ 
parallel to $\ell_1$; we get an actual triangle if and only if 
the point $q$ belongs to $S$.

In other words, our problem is now reduced to that of bounding the 
number of matching pairs in $Q$. (Since we do not enforce the condition 
that the third point $q$ of the corresponding triangle belongs to $S$, 
we most likely over-estimate the true bound.) 

Since elements of $Q$ have three degrees of freedom, we can represent
them in an appropriate 3-dimensional parametric space. For example,
we can assume that no line in $L$ is vertical, and parametrize an
element $(\ell,p)$ of $Q$ by the triple $(a,b,\kappa)$, where
$(a,b)$ are the coordinates of $p$, and $\kappa$ is the slope
of $\ell$. For simplicity of notation, we refer to this 3-dimensional
parametric space as $\reals^3$.

So far, the matching relationship is symmetric. To simplify the
analysis, and with no loss of generality, we make it assymmetric, by
requiring that, in an (ordered) matching pair
$(\ell_1,p_1)$, $(\ell_2,p_2)$, $\vec{op}_2$ lies counterclockwise to
$\vec{op}_1$, where $o=\ell_1\cap\ell_2$. See Figure~\ref{qmatch}.

Let us express the matching condition algebraically.
Let $(a,b,\kappa) \in \reals^3$ be the triple representing a pair
$(\ell,p)$, and $(x,y,w) \in \reals^3$ be the triple representing
another pair $(\ell',p')$.
Clearly, $w \not= \kappa$ in a matching pair.
The lines $\ell$ and $\ell'$ intersect at a point $o$, for which
there exist real parameters $t,s$ which satisfy
$$
o = (a+t,b+\kappa t) = (x+s,y+ws) ,
$$
or
\begin{eqnarray*}
t & = & \frac{y-b-w(x-a)}{\kappa-w} \\
s & = & \frac{y-b-\kappa(x-a)}{\kappa-w} .
\end{eqnarray*}
It is now easy to verify that the condition of matching, with
$\vec{op'}$ lying counterclockwise to $\vec{op}$, is given by
$$
\biggl( y-b-\kappa(x-a) \biggr)
\biggl( y-b-w(x-a) \biggr) = 2(w-\kappa) \quad \mbox{ and } \quad
w \not= \kappa,
$$
or, alternatively,
\begin{equation} \label{eq:m1}
w = \frac
{ \biggl(y-b-\kappa(x-a)\biggr) (y-b) + 2\kappa }
{ \biggl(y-b-\kappa(x-a)\biggr) (x-a) + 2 } \quad \mbox{ and } \quad
w \not= \kappa.
\end{equation}
Similarly, the condition of ``reverse'' matching, with $\vec{op'}$
lying clockwise to $\vec{op}$, is given by
\begin{equation} \label{eq:m1'}
w = \frac
{ \biggl(y-b-\kappa(x-a)\biggr) (y-b) - 2\kappa }
{ \biggl(y-b-\kappa(x-a)\biggr) (x-a) - 2 } \quad \mbox{ and } \quad
w \not= \kappa.
\end{equation}

Fix an element $(\ell,p)$ of $Q$, and associate with it a surface
$\sigma_{\ell,p}\subset \reals^3$, which is the locus of all
pairs $(\ell',p')$ that match $(\ell,p)$ (i.e., $(\ell,p),(\ell',p')$
is an ordered matching pair).
By the preceding analysis, $\sigma_{\ell,p}$ satisfies (\ref{eq:m1}),
where $(a,b,\kappa)$ is the parametrization of $(\ell,p)$,
and is thus a 2-dimensional algebraic surface
in $\reals^3$ of degree $3$. We thus obtain a system $\Sigma$ of 
$N$ 2-dimensional algebraic surfaces in $\reals^3$, and a set $Q$ 
of $N$ points in $\reals^3$, and our goal is to bound the number 
of incidences between $Q$ and $\Sigma$. 

The main technical step in the analysis is to rule out the possible
existence of
{\em degeneracies} in the incidence structure, where many points 
are incident to many surfaces; 
this might happen when many points lie on a common intersection
curve of many surfaces (a situation which might 
arise, e.g., in the case of planes and points in $\reals^3$). 
However, for the class of surfaces under consideration, namely, the
surfaces $\sigma_{\ell,p}$ generated by some line-point incidence pair
$(\ell,p)$, such a degeneracy is impossible, as the
following lemma shows.
\begin{lemma} \label{lem:gamma}
Let $(\ell_1,p_1)$ and $(\ell_2,p_2)$ be two distinct line-point
incidence pairs,
let $\gamma = \sigma_{\ell_1,p_1} \cap \sigma_{\ell_2,p_2}$ be the
intersection curve of their associated surfaces, and assume that
$\gamma$ is non-empty.
Let $(\ell,p)$ be some incidence pair and assume further that
$\sigma_{\ell,p} \supset \gamma$.
Then either $(\ell,p) = (\ell_1,p_1)$ or $(\ell,p) = (\ell_2,p_2)$.
\end{lemma}
\begin{proof}
We establish the equivalent claim that, given a curve $\gamma$,
which is the intersection of some unknown pair of surfaces
$\sigma_{\ell_1,p_1}$ and $\sigma_{\ell_2,p_2}$,
one can reconstruct $(\ell_1,p_1)$ and $(\ell_2,p_2)$ uniquely
(up to a swap between the two incidence pairs) from $\gamma$.
Morever, it is enough to know the projection
$\gamma^*$ of $\gamma$ onto the $xy$-plane in order to uniquely
reconstruct the incidence pairs $(\ell_1,p_1)$ and $(\ell_2,p_2)$ that
generate $\gamma$.

We start by computing the algebraic representation of $\gamma^*$.
Let $(a_1,b_1,\kappa_1)$ and $(a_2,b_2,\kappa_2)$ be the respective
parametrizations of $(\ell_1,p_1)$ and $(\ell_2,p_2)$.
By (\ref{eq:m1}), $\gamma^*$ satisfies the equation
\begin{equation} \label{eq:m12}
\frac
{ \biggl(y-b_1-\kappa_1(x-a_1)\biggr) (y-b_1) + 2\kappa_1 }
{ \biggl(y-b_1-\kappa_1(x-a_1)\biggr) (x-a_1) + 2 } =
\frac
{ \biggl(y-b_2-\kappa_2(x-a_2)\biggr) (y-b_2) + 2\kappa_2 }
{ \biggl(y-b_2-\kappa_2(x-a_2)\biggr) (x-a_2) + 2 } .
\end{equation}
Recall the additional requirement in (\ref{eq:m1}), namely that
$w \not = \kappa_1$ and $w \not = \kappa_2$. This requirement is
implicit in (\ref{eq:m1}) and in (\ref{eq:m12}), meaning that
equation (\ref{eq:m12}) is defined only for values of $x$ and $y$
for which the value of $w$ is not $\kappa_1$ or $\kappa_2$.
Consulting (\ref{eq:m1}), this implies that no point $(x,y) \in
\gamma$ can satisfy
$y-b_1 = \kappa_1(x-a_1)$ or $y-b_2 = \kappa_2(x-a_2)$.
Put
\begin{eqnarray*}
L_1 & = & y-b_1-\kappa_1(x-a_1), \quad \mbox{ and } \\
L_2 & = & y-b_2-\kappa_2(x-a_2),
\end{eqnarray*}
and write (\ref{eq:m12}) as
$$
\frac{L_1(y-b_1) + 2\kappa_1}{L_1(x-a_1)+2} =
\frac{L_2(y-b_2) + 2\kappa_2}{L_2(x-a_2)+2} ,
$$
or
$$
\big( L_1(y-b_1) + 2\kappa_1 \big)
\big( L_2(x-a_2) + 2 \big) =
\big( L_2(y-b_2) + 2\kappa_2 \big)
\big( L_1(x-a_1) + 2 \big) ,
$$
which we can rewrite as
$$
L_1L_2L_3 + 2L_1L_4 - 2L_2L_5 + 4C = 0,
$$
where
\begin{eqnarray*}
L_3 & = & (b_2-b_1)x - (a_2-a_1)y + (a_2b_1-a_1b_2) , \\
L_4 & = & y-b_1 - \kappa_2(x-a_1) , \\
L_5 & = & y-b_2 - \kappa_1(x-a_2) , \\
C & = & \kappa_1-\kappa_2 .
\end{eqnarray*}
We can further simplify the equation by noting that
$L_6 = L_1L_4 - L_2L_5$ is a linear expression is $x,y$. That is,
$$
L_6 = Dx + Ey + F ,
$$
where
\begin{eqnarray*}
D & = & 2\kappa_1\kappa_2(a_2-a_1) - (\kappa_1+\kappa_2)(b_2-b_1) , \\
E & = & 2(b_2-b_1) - (\kappa_1+\kappa_2)(a_2-a_1) , \\
F & = & \kappa_1\kappa_2(a_1^2-a_2^2)
	+ (\kappa_1+\kappa_2)(a_2b_2-a_1b_1)
	+ (b_1^2-b_2^2) .
\end{eqnarray*}
We can thus write (\ref{eq:m12}) as
\begin{equation} \label{eq:m14}
L_1L_2L_3 + 2L_6 + 4C = 0 , \quad \mbox{ and } \quad L_1 \ne 0, L_2 \ne 0.
\end{equation}
Figure \ref{fig:lines} illustrates the different lines defined by
the linear equations $L_i = 0$, and their relations with
$(\ell_1, p_1)$ and $(\ell_2,p_2)$.
The linearity of $L_1,L_2,L_3$, and $L_6$ implies that
the equation (\ref{eq:m14}) of $\gamma^*$ is cubic.
\begin{figure}[htb]
\begin{center}
\input{lines6.pstex_t} %
\caption{
	The lines $L_i = 0$, for $i = 1,\ldots,6$, in the general case.
	The line $L_3=0$
	connects $p_1$ and $p_2$, $L_4=0$ passes through $p_1$ and is
	parallel to $\ell_2$, $L_5=0$ passes through $p_2$ and is
	parallel to $\ell_1$, and $L_6=0$ connects
	$o = \ell_1 \cap \ell_2$ with the intersection point $q$ of
	$L_4=0$ and $L_5=0$ (and bisects the edge $p_1p_2$).
}
\label{fig:lines}
\end{center}
\end{figure}
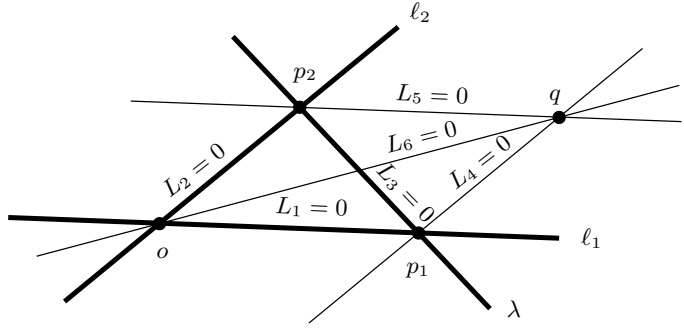
We have the following two special cases to rule out:
\begin{enumerate}
\item If $p_1 = p_2$, that is, $a_1=a_2$ and $b_1=b_2$, then
	  $L_3 = 0$, $L_4 = L_2$, and $L_5 = L_1$. But then the equation
	  becomes $4C = 0$, or $\kappa_1 = \kappa_2$, contrary to the
	  assumption that $(\ell_1,p_1) \ne (\ell_2,p_2)$. Hence, the
	  equation has no solutions, meaning that $\gamma$
	  is empty and the surfaces do not intersect.
\item If $\ell_1 = \ell_2$ but $p_1 \ne p_2$, that is,
	  $\kappa_1 = \kappa_2 = (b_2-b_1)/(a_2-a_1)$, then
	  $L_1 = L_2 = L_4 = L_5$, $L_3 = (a_1 - a_2)L_1$, and $C = 0$,
	  resulting in the equation $(L_1)^3 = 0$, which is not allowed in
	  (\ref{eq:m14}). Hence $\gamma$ is not defined in this case either.
\end{enumerate}
We can therefore restrict our attention to the general case.
Consider the cubic part of the equation $L_1L_2L_3$. In this term,
each factor can be thought of as a line defined by the equation
$L_i = 0$, for $i=1,2,3$. The lines $L_1 = 0$ and $L_2 = 0$ respectively
are simply $\ell_1$ and $\ell_2$, whereas $L_3 = 0$ is
the line $\lambda$ passing through $p_1$ and $p_2$ (see Figure
\ref{fig:lines}). Note that $\lambda$ may coincide with one of the
other two lines.
Indeed, if $p_1$ happens to be incident with $\ell_2$, then $\lambda$
coincides with $\ell_2$. Similarly, if $p_2 \in \ell_1$ then $\lambda$
coincides with $\ell_1$ (these are the only possible coincidences,
since we have ruled out the case $\ell_1 = \ell_2$).
These cases will be handled shortly, but for now, we ignore them and
consider the general
case. In this case, $\gamma^*$ has three distinct asymptotes given by
$L_1=0$, $L_2=0$, and $L_3=0$; the proof of this fact is given in Lemma
\ref{lem:as1} in the appendix

Using this fact, one can reconstruct the two line-point pairs that
generate $\gamma^*$ as follows.
Suppose we are given a curve $\gamma^*$ generated by some unknown
pair of incidence pairs, $(\ell_1, p_1)$ and $(\ell_2,p_2)$, and
we want to reconstruct these pairs.
$\gamma^*$ is given as the zero set of some cubic bivariate polynomial
$f(x,y) = 0$, where $f$ can be written as $f(x,y) =
c(L_1 L_2 L_3 + 2L_6 + 4C)$, but the decomposition of $f$
into $L_1, L_2, L_3, L_6, C$, and $c$ is unknown, and, moreover,
is not known a priori to be unique (a fact which we establish in this
proof).
First, we find its three asymptotes $\Lambda_1 = 0$, $\Lambda_2 = 0$,
and $\Lambda_3 = 0$, where for each $i=1,2,3$,
$\Lambda_i$ is linear in $x$ and $y$.
Since, by Lemma \ref{lem:as1}, these asymptotes are $L_1 = 0$, $L_2 = 0$,
and $L_3 = 0$, we know that each $\Lambda_i$ is equal to some $L_j$
multiplied by a constant, but we do not know which is which.
To determine the roles of the asymptotes correctly, observe that
$\Lambda_1 \Lambda_2 \Lambda_3 = \mu L_1 L_2 L_3$ for some constant
$\mu$. Thus, there exists some unique constant $\nu$, such that
$f(x,y) - \nu \Lambda_1 \Lambda_2 \Lambda_3 = \Lambda_4$ is linear in
$x$ and $y$. The line $\Lambda_4 = 0$ is parallel to the line $L_6 = 0$,
which happens to be the median of the triangle spanned by the three
asymptotes, which emanates from the vertex $o = \ell_1 \cap \ell_2$, and
bisects the edge $p_1p_2$; see Figure \ref{fig:lines}.
We thus have enough information to determine which vertex of the
triangle is $o$, and which are $p_1$ and $p_2$, and which edges of the
triangle are supported by $\ell_1$ and $\ell_2$.
This proves the lemma for the general case where all the points
and lines are distinct, and no point lies on both lines $\ell_2, \ell_2$.

Finally, consider the case where $p_2 \in \ell_1$ (a symmetric argument
applies when $p_1 \in \ell_2$).
In this case, $L_1 = L_5$, and $L_3 = (a_1 - a_2)L_1$, so the
equation of the curve $\gamma^*$ can be rewritten as
\[
(a_1 - a_2)L_1^2L_2 + 2L_1(L_4 - L_2) + 4C = 0.
\]
Note that $a_1 \ne a_2$ under the preliminary assumption that there
are no vertical lines in the system, since both $p_1 = (a_1,b_1)$
and $p_2 = (a_2,b_2)$ are on $\ell_1$.
Note also that $C = \kappa_1 - \kappa_2 \ne 0$, for otherwise,
$\ell_1$ and $\ell_2$ would have to coincide, a case which we
have ruled out earlier.
Finally, note that $s = L_4 - L_2 = b_2 - b_1 - \kappa_2(a_2 - a_1)
= C(a_2 - a_1)$ is a nonzero constant.
Hence, the equation of $\gamma^*$ is, up to a constant multiple,
\begin{equation} \label{eq:m13}
(a_1 - a_2)L_1^2L_2 + 2sL_1 + 4C = 0.
\end{equation}
This equation defines a cubic curve with two asymptotes given by
$L_1 = 0$, and $L_2 = 0$, namely, the lines $\ell_1$ and $\ell_2$;
the proof is given in Lemma \ref{lem:as2} in the appendix.
Since $C \ne 0$, it follows that
$\gamma^*$ does not intersect $L_1 = 0$, whereas $L_2 = 0$ is
intersected at a single point $(x,y)$ for which $L_1 = 2/(a_1-a_2)$.
Using this point, one can compute the values of $(a_1-a_2), C$, and
$s$, and hence, reconstruct the line $L_4 = 0$.
The point $p_1$ is then simply the intersection of the lines $L_1 = 0$
and $L_4 = 0$. Thus, one can uniquely reconstruct $\ell_1$, $\ell_2$,
$p_1$, and $p_2$ in this case too.
This completes the proof of Lemma \ref{lem:gamma}.
\end{proof}

\paragraph{Bounding the number of incidences.}
Recall that we need to bound the number of incidences between the set
$\Sigma$ of surfaces $\sigma_{\ell,p}$, for $(\ell,p) \in Q$,
and the set $Q$ of points. This is done by following the standard
method of Clarkson et al. \cite{CEGSW}.
The first step in this method is to derive a simple but weaker
bound, usually by extremal graph theory. Then, we strengthen the bound
by {\em cutting} the arrangement of the surfaces into cells, and by
summing the weaker bounds on the number of incidences within each cell,
over all the cells.

\paragraph{The first step: A simple bound.}
Lemma \ref{lem:gamma} implies that the incidence graph between $\Sigma$
and $Q$ does not contain $K_{3,10}$ as a subgraph, or,
in other words, no three distinct surfaces of $\Sigma$ and ten distinct
points of $Q$ can all be incident to one another. Indeed, the
intersection points of three surfaces $\sigma_{\ell_i,p_i}$, for
$i=1,2,3$, are the intersection points of the two curves
$\gamma_{1,2} = \sigma_{\ell_1,p_1} \cap \sigma_{\ell_2,p_2}$, and
$\gamma_{1,3} = \sigma_{\ell_1,p_1} \cap \sigma_{\ell_3,p_3}$. These
intersection points project to (some of) the intersection points of the
projections $\gamma^*_{1,2}$ and $\gamma^*_{1,3}$ of $\gamma_{1,2}$
and $\gamma_{1,3}$, respectively, onto the $xy$-plane.
By Lemma \ref{lem:gamma}, these two curves are distinct (or empty).
Since each of them is cubic, and since, as shown in Lemmas \ref{lem:as1}
and \ref{lem:as2} in the appendix, they are the zero sets of irreducible
polynomials,
B\'ezout's theorem \cite{Sha} implies that
they intersect in at most $3^2=9$ points.
Hence, the incidence graph between $\Sigma$ and $Q$ does not
contain $K_{3,10}$, so
by the K\H{o}vari--S\'os--Tur\'an theorem~\cite{KST54}, the
number of incidences between $\Sigma$ and $Q$ can be bounded by
\begin{equation*}
O(|\Sigma||Q|^{2/3} + |Q|).
\end{equation*}
Since the matching relation is essentially symmetric (up to some
sign changes; see (\ref{eq:m1}) and (\ref{eq:m1'})), we can interchange
the roles of points and surfaces, and conclude that the number of
incidences is also at most
\begin{equation} \label{bound:weak}
O(|Q||\Sigma|^{2/3} + |\Sigma|).
\end{equation}

\paragraph{Cutting.}
To improve the bound, we apply the following fairly standard space 
decomposition technique.
Fix a parameter $r$, whose specific value will be chosen later,
and construct a $(1/r)$-cutting $\Xi$ of $\A(\Sigma)$ \cite{Ch05}. 
We use the more simple-minded technique in which we choose a random
sample $R$ of $O(r\log r)$ surfaces of $\Sigma$ and construct the
vertical decomposition (see e.g. \cite{SA}) of the arrangement
$\A(R)$. We obtain
$O^*(r^3)$ relatively open cells of dimensions 0,1,2, and 3, each
of which is crossed by (intersected by, but not contained in) at most
$|\Sigma|/r=N/r$ surfaces; this latter property holds with high
probability, and we simply assume that our sample $R$ does satisfy it.

\paragraph{Summing over all cells.}
Fix a cell $\tau$ of $\Xi$, and put $Q_\tau := Q \cap \tau$ and $m_\tau
:= |Q_\tau|$. Let $\Sigma_\tau$ denote the subset of surfaces of $\Sigma$
which cross $\tau$, and put $N_\tau := |\Sigma_\tau| \le N/r$.

We now apply the simple bound (\ref{bound:weak}) obtained in the
first step to each cell $\tau$ of our cutting $\Xi$, handling,
for the time being, only surfaces that {\em cross} $\tau$. The
overall number of incidences is
$$
\sum_{\tau\in\Xi} O\left(
m_\tau N_\tau^{2/3} + N_\tau
\right) ,
$$
which, using the bounds $N_\tau \le N/r$, and $\sum_\tau m_\tau = N$, is
$$
O^*\left( N(N/r)^{2/3} + Nr^2 \right) =
O^*\left( N^{5/3}/r^{2/3} + Nr^2 \right) .
$$
To minimize this expression, we choose $r=N^{1/4}$, making it
$O^*(N^{3/2})$.

We also have to take into account incidences between points in a
cell $\tau$ and surfaces that fully contain $\tau$. 
This is done separately for cells of dimension $0$, $1$, and $2$ (it
is vacuous for cells of dimension $3$). Indeed, a 2-dimensional cell
$\tau$ is contained in exactly one surface, so a point $w \in \tau$
takes part in only one such incidence.
Thus, in this case we only need to add $N$, the number of points,
to the above bound.

The same argument applies for points in 1-dimensional cells.
Assuming that the vertical decomposition is performed in a generic
coordinate frame, it suffices to consider only 1-dimensional cells that
are portions of the intersection curves between the surfaces of
$\Sigma$. By Lemma \ref{lem:gamma}, each such cell $\tau$ is contained
in exactly two surfaces of $\Sigma$.
Thus, we need to add at most $2N$ to the number of incidences to handle
these cells.

Each cell of dimension $0$ is a single point $w$, and, arguing as above,
we may assume it
to be a vertex of the undecomposed arrangement $\A(R)$. Any surface
$\sigma$ incident to $w$ has to cross or bound an adjacent
full-dimensional cell $\tau^*$, so we charge the incidence of $\sigma$
with $w$ to the pair $(\tau^*,\sigma)$, and note that such a pair can
be charged only $O(1)$ times. It follows that the number of incidences
with 0-dimensional cells of $\Xi$ is $O^*(r^3+r^3(N/r)) = O^*(r^2N)$,
which, for the chosen value of $r$, is equal to the bound obtained
above for the crossing surfaces.

In conclusion, the overall number of incidences between $\Sigma$ and
$Q$ is $O^*(N^{3/2})$.

Recall now that $N=O(n^2/k^2)$, and that we also have the bound $O(n^2k)$
for the number of unit-area triangles with at least one $k$-poor
top line. Thus, the overall bound on the number of unit-area triangles
is
$$
O^*\left( \frac{n^3}{k^3} + n^2k\right) ,
$$
which, if we choose $k=n^{1/4}$, becomes $O^*(n^{9/4})$, as
asserted.
$\Box$

\paragraph{Discussion.}
Theorem \ref{thm:unit2} constitutes a major improvement over previous
bounds, but it still leaves a substantial gap from the near-quadratic
lower bound. One major weakness of our proof is that, in bounding the
number of matching pairs, it ignores the constraint that a matching
pair is relevant only when the (uniquely defined) third vertex $q$ of
the resulting triangle belongs to $S$, and that the (uniquely defined)
top line of this triangle through $q$ is $k$-rich. It is therefore
natural to conjecture that our bound is not tight, and that the true
bound is nearly quadratic, perhaps coinciding with the lower bound of
\cite{EP71}.

\appendix

\section{Asymptotes of cubic curves}
In this appendix, we analyze the class of cubic curves defined by
equations (\ref{eq:m14}) and (\ref{eq:m13}) of Section
\ref{sec:unit2}, derive their asymptotes, and show them to be the
zero sets of irreducible bivariate polynomials.
We start by analyzing a normalized version
of these equations, in which two of the generating lines
(and, as we show henceforth, the asymptotes) are the $x$ and $y$-axes.
We then reduce equation (\ref{eq:m14}) to the normalized case.
Finally, we handle equation (\ref{eq:m13}) in a different and simpler
way.

\begin{lemma} \label{lem:irr1_n}
Let $\lambda_1$ and $\lambda_2$ be two distinct lines in
$\reals^2$, given by the equations $\Lambda_i = 0$, where
$\Lambda_i = \alpha_i x + \beta_i y + \gamma_i$,
and $\alpha_i$ and $\beta_i$ are both nonzero, for $i = 1,2$.
Let $f(x,y)$ be the bivariate cubic polynomial
\begin{equation*}
f(x,y) = xy\Lambda_1 + \Lambda_2.
\end{equation*}
Then $f(x,y)$ is irreducible.
\end{lemma}
\begin{proof}
Assume, to the contrary, that $f$ is reducible. Then it has a
linear factor $L = ax + by + c$. Without loss of generality,
$b \ne 0$ (a symmetric
argument follows for the case $a \ne 0$), so we can assume $b = 1$.
Then $f$, as a polynomial in $y$ with coefficients from $\reals[x]$,
has $y = - ax - c$ as root, i.e., if we put
$$p(x) := f(x, -ax-c) = -x(ax+c)L_1 + L_2,$$
where $L_i = \alpha_i x - \beta_i (ax+c) + \gamma_i$, for $i = 1,2$,
then $p(x) \equiv 0$. But then, the term $x(ax+c)L_1$ can not be
properly cubic, nor quadratic, so, $a=c=0$, or $L_1$ is a
constant, possibly zero. In the former case, $L_2 \equiv 0$, but
$L_2 = \alpha_2x + \gamma_2$ and $\alpha_2 \ne 0$ by assumption,
a contradiction.
If $L_1 = 0$, then we
must also have $L_2 = 0$, and so both lines $\lambda_1$ and
$\lambda_2$ coincide (with the line $L = 0$), contrary to assumption.
If $L_1$ is a nonzero constant, then the term $(ax+c)$ must also
be constant, or else $p(x)$ is a proper quadratic polynomial,
hence $a=0$. But then, for $L_1 = \alpha_1 x - \beta_1 c + \gamma_1$
to be constant, we must have $\alpha_1 = 0$, in contradiction.
Either way, $f$ cannot be reducible.
\end{proof}

\begin{lemma} \label{lem:as_n}
Let $\lambda_1$ and $\lambda_2$ be two distinct lines in $\reals^2$,
given by the equations $\Lambda_i = 0$, where
$\Lambda_i = \alpha_i x + \beta_i y + \gamma_i$, for $i = 1,2$,
such that $\alpha_1$ and $\beta_1$ are both nonzero.
Let $\Gamma$ be the algebraic cubic curve defined by the equation
\begin{equation} \label{eq:n14}
xy\Lambda_1 + \Lambda_2 = 0.
\end{equation}
Then $\Gamma$ is asymptotic to the $x$-axis and to the $y$-axis.
\end{lemma}
\begin{proof}
We only prove in detail that the $x$-axis is an asymptote. Note
that, for any fixed $x \ne 0$, (\ref{eq:n14}) is a quadratic
equation in $y$, which we rewrite as
\[
xy(\alpha_1 x + \beta_1 y + \gamma_1)
	+ \alpha_2 x + \beta_2 y + \gamma_2 = 0,
\]
or
\[
\beta_1 xy^2 + (\alpha_1 x^2 + \gamma_1 x + \beta_2)y
	+ (\alpha_2 x + \gamma_2) = 0.
\]
Hence
\begin{eqnarray*}
y & = & - \frac{\alpha_1 x^2 + \gamma_1 x + \beta_2}{2\beta_1 x} \\
  &   & \pm \frac{\sqrt{(\alpha_1 x^2 + \gamma_1 x + \beta_2)^2
		  		    - 4\beta_1 x(\alpha_2 x + \gamma_2)}}
		 {2\beta_1 x}.
\end{eqnarray*}
We only consider the solution with positive square root, which is
\[
y = \frac{\alpha_1 x^2 + \gamma_1 x + \beta_2}{2\beta_1 x}
	\left[
		\sqrt{1 - \frac{4\beta_1 x(\alpha_2 x + \gamma_2)}
					   {(\alpha_1 x^2 + \gamma_1 x + \beta_2)^2}}
		- 1
	\right].
\]
The expression in the square brackets is of the form
$\sqrt{1+t}-1$. Since $\alpha_1 \ne 0$, $t$ tends to 0 as $x \to
\pm \infty$. Using the inequalities
$1 - |t| \le \sqrt{1+t} \le 1 + \frac{|t|}{2}$,
for $|t| < 1$, we obtain, for $|x|$ sufficiently large,
\[
|y| \le \frac{|\alpha_1 x^2 + \gamma_1 x + \beta_2|}{|2\beta_1 x|}|t|
	= \frac{2|\alpha_2 x + \gamma_2|}{|\alpha_1 x^2 + \gamma_1 x + \beta_2|},
\]
which tends to 0 as $x \to \pm \infty$. This shows that the
$x$-axis is indeed an asymptote of $\Gamma$ (on both sides).
A symmetric argument shows that the $y$-axis is also an
asymptote.
\end{proof}

We are now ready to prove the more general cases discussed in
Section \ref{sec:unit2}.
\begin{lemma} \label{lem:as1}
Let $\ell_1, \ldots, \ell_4$ be four distinct lines in
$\reals^2$, given by the equations $L_i = 0$, where
$L_i = A_i x + B_i y + C_i$, for $i = 1,\ldots,4$.
Assume that no pair of $\ell_1, \ell_2, \ell_3$ are parallel,
and that $\ell_4$ is not parallel to any of $\ell_1$ and $\ell_2$.
Put
$$f(x,y) = L_1 L_2 L_3 + L_4,$$
and
let $\Gamma$ be the algebraic cubic curve defined by the equation
\[
f(x,y) = 0.
\]
Then, $f$ is irreducible, and $\Gamma$ is asymptotic to the lines
$\ell_1, \ell_2, \ell_3$.
\end{lemma}
\begin{proof}
We may assume, by an appropriate change of variables, that one of
$\ell_1, \ell_2$, and $\ell_3$ is the $x$-axis and another one is
the $y$-axis. For example, put $u = L_1$, and $v = L_2$, and write
$L_3 = \alpha_1 u + \beta_1 v + \gamma_1$, and
$L_4 = \alpha_2 u + \beta_2 v + \gamma_2$,
for some appropriate coefficients
$\alpha_1, \beta_1, \gamma_1, \alpha_2, \beta_2, \gamma_2$.
Note that, by the preliminary assumptions on the lines,
$\alpha_i$ and $\beta_i$ are both nonzero, for $i=1,2$.
$\Gamma$ can then be written as
\[
g(u,v) = uv L_3 + L_4 = 0
\]
in the $(u,v)$ coordinate system. It then follows, by Lemma
\ref{lem:irr1_n}, that $f$ is irreducible, for otherwise, any
factorization of $f$ could be transformed into a factorization of
$g$, in contradiction.
It also follows, by Lemma \ref{lem:as_n}, that $\ell_1$ and
$\ell_2$ are asymptotes of $\Gamma$.
Note that, for this part of the argument, the choice of 
$\ell_1$ and $\ell_2$ as axes is arbitrary, and we could just 
as well choose any other pair of lines among $\ell_1,\ell_2,\ell_3$.
In more detail, since no pair of these three lines are parallel, we
can make any two of them as the axes of a new $(u,v)$-coordinate 
system, and then, in the equation of the third line, both the $u$- and
$v$-coefficients would be nonzero, which is the condition assumed in 
Lemma \ref{lem:as_n}. Hence, $\ell_3$ is also an asymptote of $\Gamma$.
\end{proof}

\begin{lemma} \label{lem:as2}
Let $\ell_1$ and $\ell_2$ be two distinct intersecting lines in
$\reals^2$, given by the equations $L_i = 0$, where
$L_i = A_i x + B_i y + C_i$, for $i = 1,2$.
Put $f(x,y) = L_1^2 L_2 + L_1 + C$, for some constant $C$, and
let $\Gamma$ be the algebraic curve defined by the equation
\[
f(x,y) = 0.
\]
Then $\Gamma$ is asymptotic to the lines $\ell_1$ and $\ell_2$.
Furthermore, if $C \ne 0$, then $f$ is an irreducible bivariate
polynomial.
\end{lemma}
\begin{proof}
If $C = 0$, then the claim is easy. Indeed, in this case we
have $L_1(L_1L_2 + 1) = 0$, so $\Gamma$ is the union of the line
$L_1 = 0$ and the hyperbola $L_1L_2 = -1$, which is asymptotic to
the lines $L_1 = 0$, and $L_2 = 0$.

If $C \ne 0$, put $u = L_1$, and $v = L_2$. Then, in the $(u,v)$
coordinate system, $\Gamma$ is defined by the equation
\[
g(u,v) := u^2v + u + C = 0.
\]
Note that $g$ is clearly irreducible, and so is $f$.
This equation can be rewritten as
\[
v = -\frac{u+C}{u^2}.
\]
Clearly, this function tends to 0 as $u$ tends to $\infty$, which
means that $\Gamma$ is asymptotic to the $u$-axis, i.e.,
to $\ell_2$.
Furthermore, the function has a pole at $u = 0$, meaning that
$\Gamma$ is asymptotic to the $v$-axis, i.e., to $\ell_1$.
\end{proof}
\end{document}

%% file: qmatch.pstex_t
\begin{picture}(0,0)%
\includegraphics{qmatch.pstex}%
\end{picture}%
\setlength{\unitlength}{3947sp}%
\begingroup\makeatletter\ifx\SetFigFontNFSS\undefined%
\gdef\SetFigFontNFSS#1#2#3#4#5{%
  \reset@font\fontsize{#1}{#2pt}%
  \fontfamily{#3}\fontseries{#4}\fontshape{#5}%
  \selectfont}%
\fi\endgroup%
\begin{picture}(3617,3179)(730,-3238)
\put(4253,-2520){\makebox(0,0)[lb]{\smash{{\SetFigFontNFSS{12}{14.4}{\rmdefault}{\mddefault}{\updefault}$\ell_1$}}}}
\put(2303,-218){\makebox(0,0)[lb]{\smash{{\SetFigFontNFSS{12}{14.4}{\rmdefault}{\mddefault}{\updefault}$\ell_2$}}}}
\put(3098,-1021){\makebox(0,0)[lb]{\smash{{\SetFigFontNFSS{12}{14.4}{\rmdefault}{\mddefault}{\updefault}$q$}}}}
\put(2333,-2513){\makebox(0,0)[lb]{\smash{{\SetFigFontNFSS{12}{14.4}{\rmdefault}{\mddefault}{\updefault}$p_1$}}}}
\put(1651,-1036){\makebox(0,0)[lb]{\smash{{\SetFigFontNFSS{12}{14.4}{\rmdefault}{\mddefault}{\updefault}$p_2$}}}}
\put(1808,-1898){\makebox(0,0)[lb]{\smash{{\SetFigFontNFSS{12}{14.4}{\rmdefault}{\mddefault}{\updefault}{\color[rgb]{0,0,0}$1$}%
}}}}
\put(2348,-1509){\makebox(0,0)[lb]{\smash{{\SetFigFontNFSS{12}{14.4}{\rmdefault}{\mddefault}{\updefault}{\color[rgb]{0,0,0}$1$}%
}}}}
\end{picture}%

%% file: lines6.pstex_t
\begin{picture}(0,0)%
\includegraphics{lines6.pstex}%
\end{picture}%
\setlength{\unitlength}{4144sp}%
\begingroup\makeatletter\ifx\SetFigFontNFSS\undefined%
\gdef\SetFigFontNFSS#1#2#3#4#5{%
  \reset@font\fontsize{#1}{#2pt}%
  \fontfamily{#3}\fontseries{#4}\fontshape{#5}%
  \selectfont}%
\fi\endgroup%
\begin{picture}(4073,1984)(170,-1280)
\put(3694,-793){\makebox(0,0)[b]{\smash{{\SetFigFontNFSS{9}{10.8}{\rmdefault}{\mddefault}{\updefault}{\color[rgb]{0,0,0}$\ell_1$}%
}}}}
\put(3475, 73){\makebox(0,0)[b]{\smash{{\SetFigFontNFSS{9}{10.8}{\rmdefault}{\mddefault}{\updefault}{\color[rgb]{0,0,0}$q$}%
}}}}
\put(3231,-1225){\makebox(0,0)[b]{\smash{{\SetFigFontNFSS{9}{10.8}{\rmdefault}{\mddefault}{\updefault}{\color[rgb]{0,0,0}$\lambda$}%
}}}}
\put(2476,-241){\rotatebox{14.0}{\makebox(0,0)[lb]{\smash{{\SetFigFontNFSS{9}{10.8}{\rmdefault}{\mddefault}{\updefault}{\color[rgb]{0,0,0}$L_6 = 0$}%
}}}}}
\put(2521, 74){\rotatebox{357.5}{\makebox(0,0)[lb]{\smash{{\SetFigFontNFSS{9}{10.8}{\rmdefault}{\mddefault}{\updefault}{\color[rgb]{0,0,0}$L_5 = 0$}%
}}}}}
\put(1171,-511){\rotatebox{38.0}{\makebox(0,0)[lb]{\smash{{\SetFigFontNFSS{9}{10.8}{\rmdefault}{\mddefault}{\updefault}{\color[rgb]{0,0,0}$L_2 = 0$}%
}}}}}
\put(1801,-601){\rotatebox{357.5}{\makebox(0,0)[lb]{\smash{{\SetFigFontNFSS{9}{10.8}{\rmdefault}{\mddefault}{\updefault}{\color[rgb]{0,0,0}$L_1 = 0$}%
}}}}}
\put(2670,557){\makebox(0,0)[b]{\smash{{\SetFigFontNFSS{9}{10.8}{\rmdefault}{\mddefault}{\updefault}{\color[rgb]{0,0,0}$\ell_2$}%
}}}}
\put(2881,-466){\rotatebox{38.0}{\makebox(0,0)[lb]{\smash{{\SetFigFontNFSS{9}{10.8}{\rmdefault}{\mddefault}{\updefault}{\color[rgb]{0,0,0}$L_4 = 0$}%
}}}}}
\put(2656,-961){\makebox(0,0)[b]{\smash{{\SetFigFontNFSS{9}{10.8}{\rmdefault}{\mddefault}{\updefault}{\color[rgb]{0,0,0}$p_1$}%
}}}}
\put(1126,-871){\makebox(0,0)[b]{\smash{{\SetFigFontNFSS{9}{10.8}{\rmdefault}{\mddefault}{\updefault}{\color[rgb]{0,0,0}$o$}%
}}}}
\put(1981,209){\makebox(0,0)[b]{\smash{{\SetFigFontNFSS{9}{10.8}{\rmdefault}{\mddefault}{\updefault}{\color[rgb]{0,0,0}$p_2$}%
}}}}
\put(2400,-354){\rotatebox{314.0}{\makebox(0,0)[lb]{\smash{{\SetFigFontNFSS{9}{10.8}{\rmdefault}{\mddefault}{\updefault}{\color[rgb]{0,0,0}$L_3 = 0$}%
}}}}}
\end{picture}%